
\input harvmac

\input epsf
\epsfverbosetrue

\Title{BROWN-HET-943, hep-ph/9407299}{High Energy Scattering in 2+1 QCD:
A Dipole Picture }
\centerline{Miao Li\foot{E-mail: li@het.brown.edu}
and Chung-I Tan\foot{E-mail: tan@het.brown.edu}}
\bigskip
\centerline{\it Department of Physics}
\centerline{\it Brown University}
\centerline{\it Providence, RI 02912}
\bigskip

A dipole picture of high energy scattering is developed in the 2+1
dimensional QCD, following Mueller. A generalized integral equation for the
dipole density with a given separation and center of mass position
is derived, and meson-meson
non-forward scattering amplitude is therefore calculated. We also calculate the
amplitude due to two pomeron exchange, and the triple pomeron coupling. We
compare the result obtained
by this method to our previous result based on an effective action
approach, and find the two results agree at the one pomeron exchange level.

\Date{7/94}

\newsec{Introduction}
One of
the most striking aspects of high-energy hadron-hadron scattering is the
continued increase of the total cross section
$\sigma_T$ with the energy. Traditional approaches to high energy near-forward
hadronic collisions invariably involve the notion of a Pomeron exchange.
There are currently two
seemingly conflicting interpretations of Pomeron in QCD: One  based on
perturbative leading log approximation (LLA)
\ref\Lipatov{L.N. Lipatov, review in {\it Perturbative QCD}, ed. A.H.
Mueller (World Scientific, Singapore, 1989), and references therein.}
\ref\Leningrad{L.V. Gribov, E.M. Levin, and M.G. Ryskin, Phys. Rep.
  100C (1983) 1; E.M. Levin and M.G. Ryskin, Phys. Rep.  189C (1990)
267.}
and another  based on nonperturbative (large-$N_c$ and/or phenomenological)
considerations \ref\DPM{A. Capella, U. Sukhatme,
C-I Tan and Tran T. V., Phys. Lett.  B81 (1979) 68;
see also: {A. Capella, U. Sukhatme,
C-I Tan and Tran T. V., Phys. Rep. 236 (1994) 225.}} \ref\Landshoff{P. V.
Landshoff, Proc. of 3rd
Int. Conference on Elastic and Diffractive Scattering, Nucl. Phys.  B12
(1990) 397.}. Thus one of the main puzzles for our understanding of high
energy hadronic collisions
in QCD is the relation, if any, between the perturbative (Lipatov) Pomeron
and the nonperturbative (soft) Pomeron.
In a perturbative treatment, a Pomeron loosely corresponds to
the color-singlet bound state of  two (reggeized) gluons.
In a
nonperturbative treatment, the Pomeron is thought  to correspond to
nonperturbative gluon exchanges having the topology of a closed string.
It has been emphasized in \ref\tan{E. M. Levin and C-I Tan, ``Heterotic
Pomeron", Proc. of XXII
Int. Symposium on Multiparticle Dynamics, Santiego, Spain, (World Scientific,
Singapore, 1993).}
 that  a ``two-gluon ladder" also  has the  topological structure of a
``cylinder" in the color
space. This fact offers  the possibility of a
unified approach to both perturbative and nonperturbative Pomerons in a
systematic large-$N_c$ treatment.

Two recent studies in  QCD at high energies, together with the small-$x$
results coming from HERA, have heightened interests in this area of research.
Formally,  a high energy
hadronic collision in the near-forward limit corresponds to the mixing of
a ``short-distance"
phenomenon in the longitudinal coordinates with  a ``long-distance"
phenomenon in the transverse
coordinates. By treating the longitudinal and transverse degrees of freedom
separately, one could
hope to achieve   a ``dimensional reduction", simplifying QCD at high energies
to an effective two-dimensional field theory.\ref\goal{This indeed has always
been the goal of
both the traditional multi-peripheral approach to high energy scattering and
the LLA of Cheng-Wu,
Lipatov, {\it etc.}. See Ref. [12] by Horn and Zachariasen, and H. Cheng and
T.T.
Wu, Expanding Protons: Scattering at High Energies, The MIT Press (1987).}  An
interesting attempt in this
direction has been made recently by Verlinde and Verlinde
\ref\vv{H. Verlinde and E. Verlinde, preprint PUPT-1319, revised version.}.
The effective
theory involves several fields and two coupling constants, the original gauge
coupling
$g$ and an effective coupling $e^2 \sim g^2\log s$. Like  previous attempts
by Lipatov and
collaborators, the emphasis is on first  considering  individual quark-quark
scattering through which one reconstructs physical hadronic scattering
amplitudes.

The second important development is due to Mueller\ref\am{A.H. Mueller, Nucl.
Phys. B415 (1994) 373.}\ref\mp{
A.H. Mueller and B. Patel, Columbia preprint, March (1994).}, who provided an
intuitively attractive  dipole picture for the high energy hard processes.
 In this approach,
which is based on a large-$N_c$ analysis, the basic scattering
 is between a pair of color dipoles through a two-gluon exchange mechanism and
the  whole
complexity of high energy hadronic collisions lies in calculating the wave
function and the ``dipole
density" in the transverse space created by  a fast moving hadron.
 Interestingly, the
BFKL pomeron is rederived at the stage of calculating the dipole density. For
related work, see
\ref\russian{ N.N. Nikolaev, B.G. Zakharov and V.R. Zoller,
KFA-IKP(Th)-1993-34, November (1993).}. One possible advantage of this new
approach is the relative ease in
deriving the BFKL pomeron and generalizing it to multi-pomerons case.

In this paper, we carry out a study  for QCD in 2+1 dimensions using this
dipole picture. One of the
purposes of this study is to clarify as well as to amplify the large-$N_c$
cylinder structure, {\it
i.e.,} closed-string like structure,  of the Lipatov Pomeron. Since much
simpler analytic
calculations are involved, the resulting physical picture becomes clearer.
Secondly,  since we have
previously carried out a similar analysis for 2+1 QCD using Verlindes'
approach\ref\lt{M. Li and C-I
Tan, Phys. Rev. D, in press.}, this study also helps to check on  the validity
of this alternative
approach.  One intriguing result of Ref.\lt\ is the fact that the total cross
section due to one-Pomeron exchange does not grow with energy.
We shall reproduce this result using this dipole approach and identify its
physical origin as due to color-screening in 2+1 dimensions. Even more
importantly, this study potentially could help to
clarify and contrast the relation of a perturbative approach to that of a
nonperturbative soft Pomeron, which also relies on a large-$N_c$ expansion.

 In Sec. 2 we first derive an
integral equation governing the light-cone meson wave function with emission
of soft gluons. A physical interpretation for the
virtual correction factor involved in the integral equation is also provided.
We emphasize the
surface interpretation of the dipole picture in Sec. 3, thus providing  a
possible link
between this perturbative scheme and other  large $N_c$ nonperturbative
schemes. The integral
equation for the dipole density is solved in Sec. 3, and the meson-meson
scattering amplitude due to one Pomeron exchange is calculated and found,
in agreement with results of Ref. \lt. We   proceed in Sec. 4 to derive and
then solve
 the integral equation for dipole pair density. A surprisingly simple
result is obtained which allows one to calculate the contribution to an elastic
amplitude due to
  two Pomeron exchange. The triple Pomeron
coupling is discussed in Sec. 5. A brief discussion of our results is given in
Sec. 6.

\newsec{The meson wave function}

In this section we shall write down the integral equation governing the
generating functional of probability amplitudes of soft gluons in a given
fast moving meson in 2+1 dimensions. This is a preparation for deriving an
integral
equation governing the dipole density in a meson state. We  adopt the approach
of \am\ and try to
stay as close as possible with notations introduced in \mp.

Let $\psi(k,z)$ be the light-cone wave function of the meson containing
no soft gluons, where $k$ is the relative transverse momentum between
quark and anti-quark and there is only one component in our situation,
$z$ is the fraction of total longitudinal momentum
carried by quark. We shall ignore spinor indices, since they are irrelevant in
the following derivation. The light-cone wave function with one soft
gluon is
\eqn\glu{\psi^a(k_1,k_2,z_1,z_2)=2gT_a\left(\psi(k_1,z_1)-\psi(k_1+k_2,z_1)
\right){k_2\over k_2^2+\mu^2},}
where $k_1$ is the transverse momentum carried by quark, $k_2$ is the
transverse momentum carried by the soft gluon, $z_2$ is the fraction of
total longitudinal momentum carried by this soft gluon and is assumed to be
much smaller
than $z_1$. The formula is nearly identical to that obtained in \am\ and is
schematically
represented by  diagrams in figure 1. Note that in 2+1 dimensions, there is
only one polarization and we assumed the polarization constant be unity in the
above formula. A
infrared cut-off $\mu$ is also introduced.
\vskip1cm
\epsfxsize=300pt \epsfbox{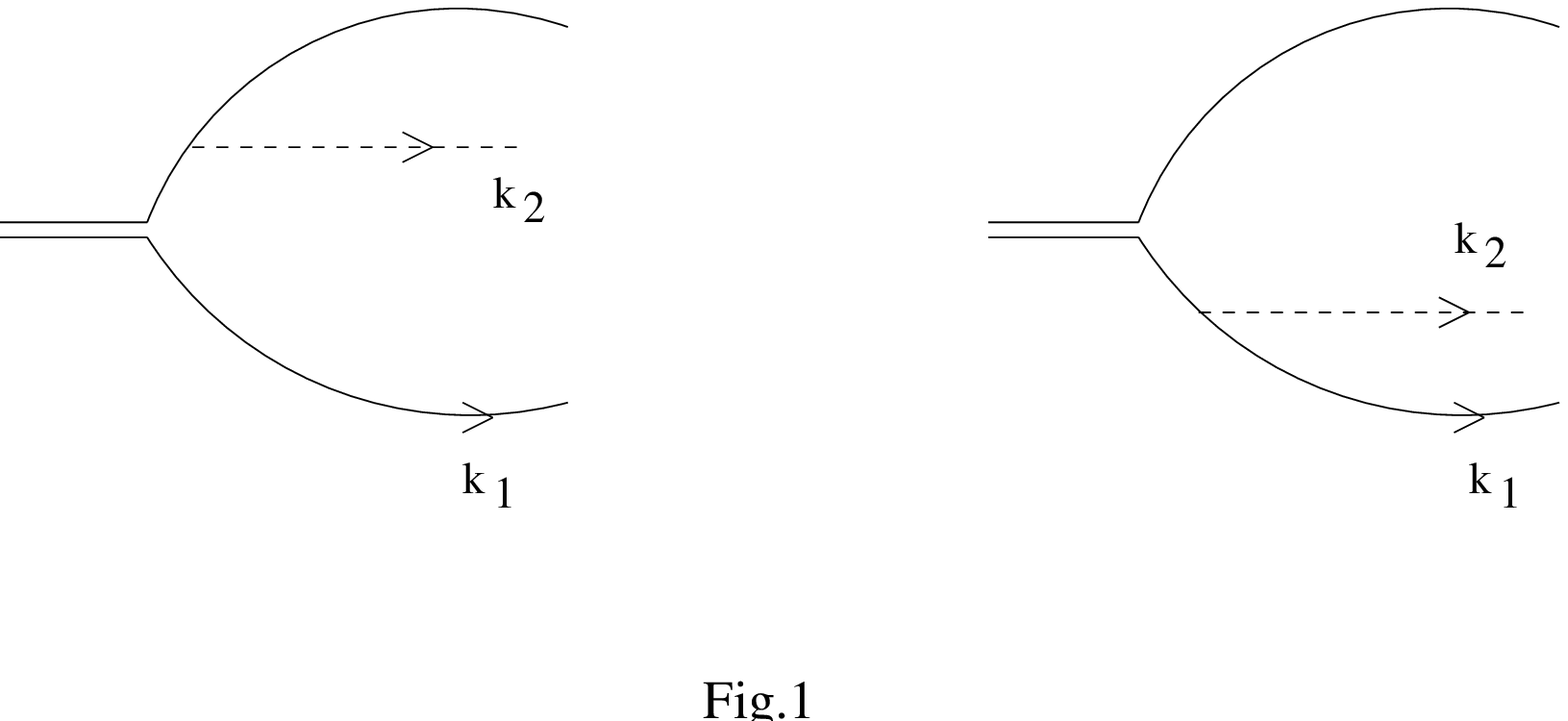}

It proves convenient to express
the expression \glu\ in coordinate space,
\eqn\sglu{\psi^a(x_0,x_1,x_2,z_1,z_2)=igT_a\psi(x_0,x_1,z_1)[\epsilon(x_2-x_0)
-\epsilon(x_2-x_1)],}
where $\epsilon$ is the step function, {\it i.e.}, $\epsilon(x)=1 $ if $x>0$
and  $\epsilon(x)=-1$  if
$x<0$.  Just as in the 3+1 dimensional case, the wave function with one soft
gluon factorizes:  It is the product of the
wave function of valence quarks and an elementary function.  It follows that
the probability
for  one soft gluon emission is then
\eqn\prob{\phi^{(1)}(x_0,x_1,x_2,z_1,z_2)={g^2\over\pi}C_F K(x_{20},x_{21})
\phi^{(0)}(x_0,x_1,z_1),}
where $C_F\equiv 1/2 (N_c-1/N_c)\rightarrow N_c/2$ in the large $N_c$ limit,
the factor $1/\pi$
comes from the longitudinal integration measure, and
\eqn\kernel{ K(x_{20},x_{21})=(1/4 )[\epsilon(x_{20})-\epsilon(x_{21})]^2.}
However, unlike the situation of 3+1 dimensions, the kernel $K(x_{20},x_{21})$
vanishes for $x_2$
outside of the interval $(x_0,x_1)$, (we assume that $x_1>x_0$). This  special
property of one
dimensional propagator implies that,  due to color screening, no gluon can be
emitted outside of the
``parent" dipole in the transverse impact space. This fact is ultimately
responsible for  our final
result that the total cross section does not grow with energy.

Similarly, the probability  of finding two
soft gluons in the meson is
\eqn\twglu{\phi^{(2)}(x_i,z_i)=\lambda(K(x_{32},x_{30})
+K(x_{31},x_{32}))\lambda K(x_{20},x_{21})\phi^{(0)}(x_0,x_1,z_1),}
where $\lambda=g^2N_c/(2\pi)$.
Since in the large-$N_c$ limit the color structure of a gluon can be identified
with that of
a quark-antiquark pair, \twglu\ has a simple interpretation in  a dipole
picture: The  second gluon
at
$x_3$ is either emitted from the dipole consisting of one valence quark and the
anti-quark part in
the first gluon, or emitted from the dipole consisting of the valence
anti-quark and the quark part
of the first gluon. We emphasize  that one could speak of splitting a gluon
into a pair of quark
and anti-quark only in a  large-$N_c$ language. In this next section, a clearer
discussion on the relation of this picture to the usual large-$N_c$ planar
graph analysis will be provided.

As in \am, we next introduce a generating functional $\Phi(x_0,x_1,z_1,
u(x,z))$
whose $n$th-order coefficient in an expansion in $u(x,z)$ is the probability
amplitude of
finding $n$ soft gluons in the meson. Clearly, $\Phi$ factorizes into
$\phi^{(0)}
(x_0,x_1,z_1)Z(x_0,x_1,u(x,z))$.  One can of course obtain  these
coefficients iteratively. More directly, it follows formally from
\prob\ and
\twglu\ that $Z$ satisfies the following integral equation:
\eqn\nonl{Z(x_0,x_1,z_1,u)=1+\lambda\int_{x_0}^{x_1}dx_2\int_{z_0}^{z_1}
{dz_2\over z_2}u(x_2,z_2)Z(x_2,x_1,z_2,u)Z(x_0,x_2,z_2,u).}
The structure of this integral equation can also be seen more clearly by
examining figure 2. Note
that a cut-off
$z_0$ to the longitudinal integration over
$z_2$ has been introduced.
\vskip1cm
\epsfxsize=320pt \epsfbox{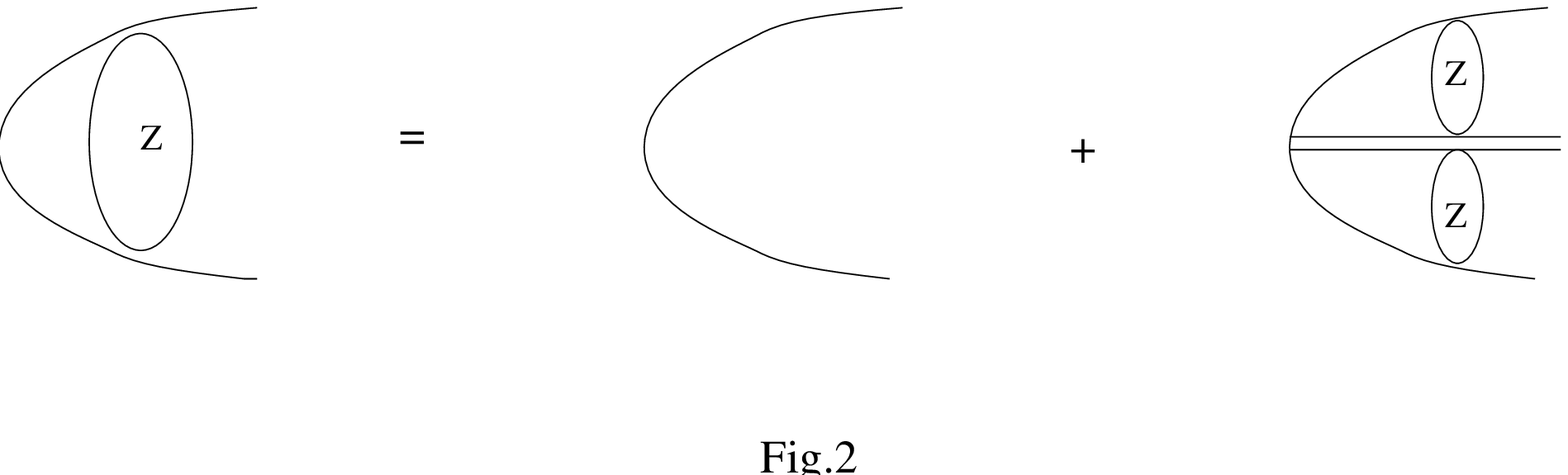}

 However, \nonl\ as it stands is inconsistent, as pointed out in \am\ for the
analogous equation
in the case of  3+1 dimensions.  If we define $\phi^{(0)}(x_0,x_1,z_1)$ as the
inclusive
probability of finding the  valence quark and anti-quark in the meson, then
production of soft gluons
should not change the probability. Consequently one has
$\phi^{(0)}(x_0,x_1,z_1)=\Phi(x_0,x_1,
z_1,u=1)$, or, $Z(x_0,x_1,z_1,u=1)=1$. However, this condition is not
satisfied by \nonl. To rectify the situation, we notice that the ``bare''
wave function $\phi^{(0)}$ used in \prob\ must be replaced by $\phi^{(0)}
A(x_0,x_1,z_1)$ where the factor  $A$ takes into account  virtual
corrections to the emission of one real gluon. A similar factor must be used
in each dipole
constructed from emitted gluons. Since
$Z(x_0,x_1,z_1,0)=A(x_0,x_1,z_1)$,  it follows that it should appear as the
inhomogeneous term in
\nonl. Introducing rapidities $y=\ln (z_1/z_0)$, $y'=\ln (z_2/z_0)$, it is
convenient to treat  the
soft gluon emission as an evolution process. Normalizing $A(x_0,x_1,z_0)=1$, it
can be interpreted
as the survival probability of no soft gluon emission at $y>0$. It  follows
that \nonl\ should
be further modified by inserting a factor $A(x_0,x_1,z_1z_0/z_2)$, i.e., the
probability of no gluon
emission as one evolves from $y'$ to $y$.

This survival probability function satisfies a first order differential
equation $dA/dy=-\lambda\int dx{_2}K(x_{20},x_{21})A$. It follows that
$A=\exp(-\lambda
x_{10}y)$. The integral equation \nonl\ is  thus modified to become
\eqn\modi{\eqalign{Z(x_0,x_1,y,u(x,y))=&e^{-\lambda x_{10}y}+
\lambda\int_{x_0}^{ x_1}dx_2\int_0^ye^{-\lambda x_{10}(y-y')}dy'u(x,y')\cr
&Z(x_2,x_1,y',u)Z(x_0,x_2,
y',u).}}
 Note that the desired normalization condition at $u=1$ is
satisfied by \modi.
There is no ultraviolet divergence in the virtual corrections in 2+1 dimensions
since  QCD is
super-renormalizable, unlike the situation for 3+1 dimensions.

\newsec{One Pomeron exchange}

\subsec{Topological structure of a meson state}

Knowing the integral equation for the generating functional $Z$, it is
straightforward to write down the integral equation for the dipole density.
Before doing so, let us first provide an intuitive picture of such a density
function from the
point of view of large-$N_c$  expansion where the dominant topological
structure is
that of closed-string exchange, figure 3.
The cylindrical surface in figure 3 is understood to
have been populated with arbitrary number of gluons, provided that they are
connected in a
``planar" fashion in their color structure. This can best be represented if
each gluon is
labeled in a ``double-line" representation, {\it i.e.}, as far as color is
concerned, a gluon can
be thought of as a pair of quark and antiquark.  In this fashion, the
cylindrical surface is
``tiled" by polygons whose edges are gluon lines. The color structure is such
that the boundary
of each polygon corresponds to  a closed ``quark loop", formed by identifying a
quark line from each
gluon of the polygon edges.
\vskip1cm
\epsfxsize=250pt \epsfbox{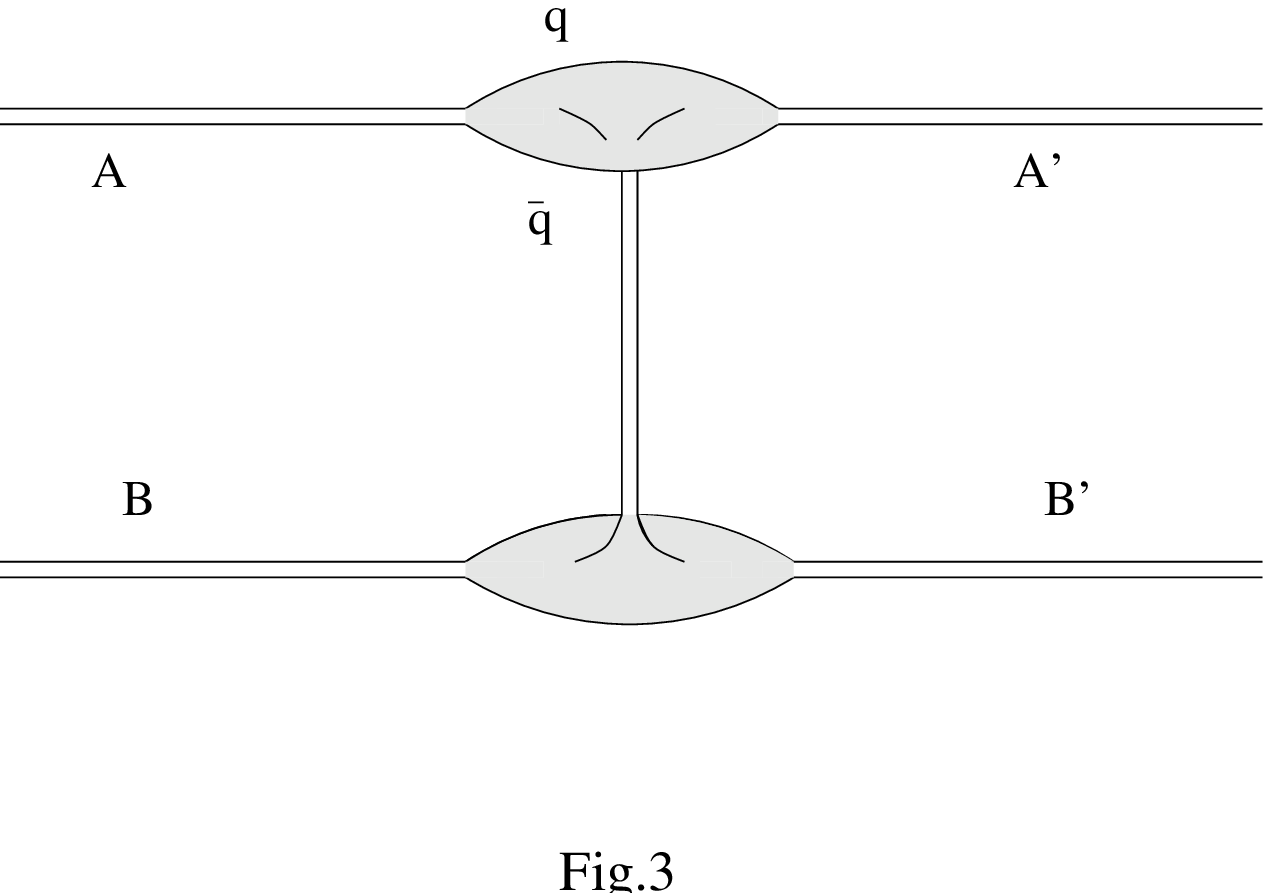}

Figure 3 can also be obtained by  joining of a closed-string propagator, {\it
i.e.}, the long
neck, to wave functions at both ends of the graph. By construction, this
closed-string
propagator corresponds to a color-neutral exchange. (In what follows, it will
be represented by
the color-neutral component of two-gluon exchange. Here we shall concentrate on
the topological
structure of the  wave-functions at each end.)   Each wave function, before the
closed-string
exchange, topologically has the structure of a disk. In order to join to the
closed-string, a
``window", or more precisely, a ``quark-line" boundary must be cut out. In a
light-cone wave
function approach, this quark-loop represents the propagation of a
color-dipole.  The ``distribution" of this quark-loop  can then be understood
as the desired dipole density. The quark-gluon diagram and the corresponding
surface diagram are illustrated in figure 4.
\vskip1cm
\epsfxsize=240pt \epsfbox{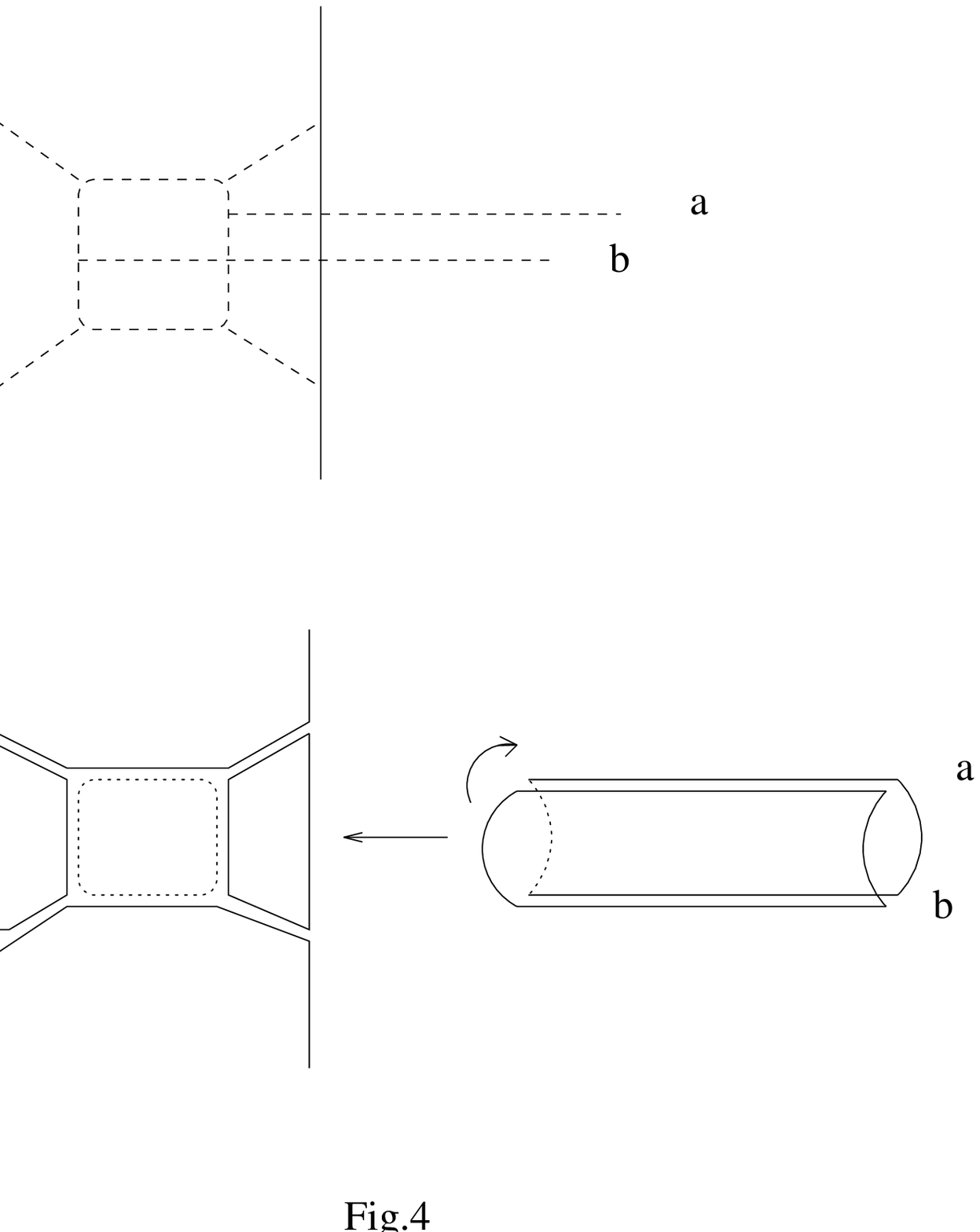}

\subsec{The dipole density in a meson state}

Let us now return to our treatment of the light-cone wave function of a fast
meson. Let $n(x_0,x_1,x,\bar{x},y)$ be the dipole density in a meson with
valence quark at $x_0$ and anti-quark
at $x_1$. The dipole has a separation
$x$ and center of mass position $\bar{x}$.  The density is defined for
dipoles with
the lowest longitudinal momentum greater than
$\exp(-y)p_+$, where $p_+$ is  the total longitudinal momentum of the meson. By
keeping track of the center of mass, our treatment
here represents an extension of that treated in \am\ and \mp.

{}From graphs in figure 5, it is easy to see that the integral equation
governing the dipole density is
\eqn\dd{\eqalign{n(x_0,x_1,x,\bar{x},y)=&e^{-\lambda x_{10}y}\delta(|x|-
x_{10})\delta(\bar{x}-\bar{x}_{10})\cr
&+\lambda\int_{x_0}^{x_1}dx_2\int_0^ydy'e^{-\lambda x_{10}(y-y')}[n(x_0,x_2,x,
\bar{x},y')+n(x_2,x_1,x,\bar{x},y')],}}
where $\bar{x}_{10}$ is the center of mass position $1/2(x_0+x_1)$ of valence
quarks. The delta function in the first term represents the contribution
from the valence quarks. Again we assume $x_1>x_0$.
\vskip1cm
\epsfxsize=340pt \epsfbox{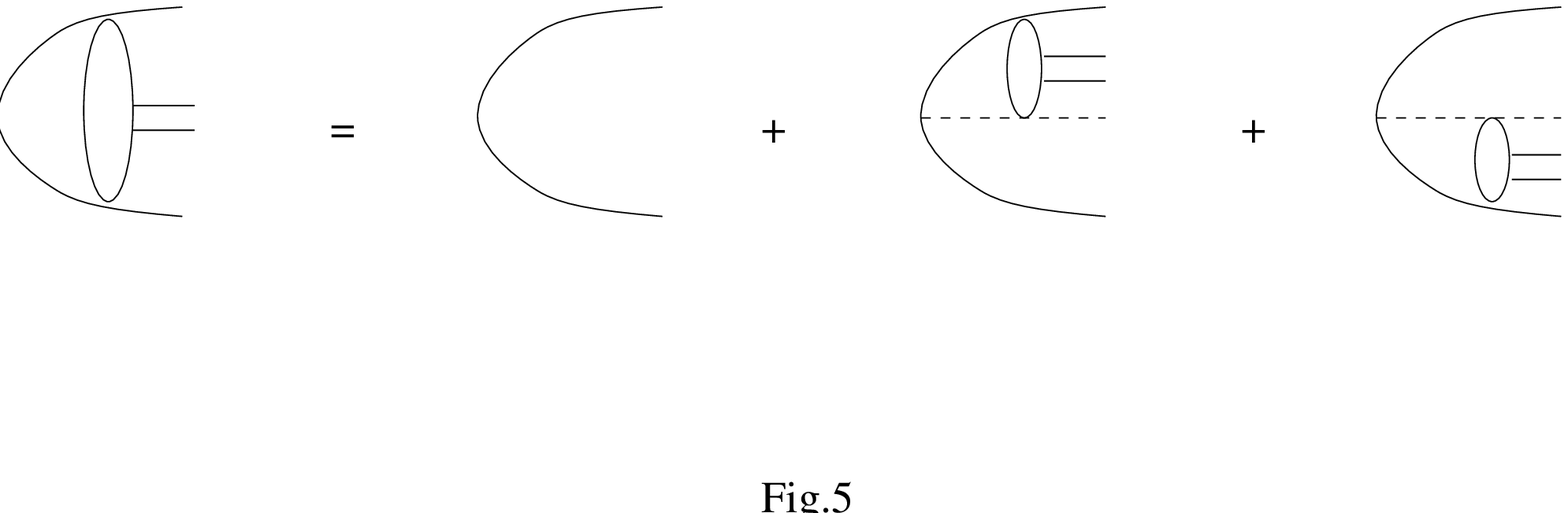}

To solve Eq. \dd, use
\eqn\trans{n(x_0,x_1,x,\bar{x},y)=\int{d\omega\over 2\pi i}e^{y\omega}
n_\omega(x_0,x_1,x,\bar{x}),}
where the contour of the integral runs parallel to the imaginary axis and
to the right of all singularities in $n_\omega$. The integral equation
for $n_\omega$ becomes
\eqn\ntrans{\eqalign{(\omega+\lambda x_{10})n_\omega(x_0,x_1,x,\bar{x})=&\delta
(|x|-x_{10})\delta(\bar{x}-\bar{x}_{10})\cr
&+\lambda\int_{x_0}^{x_1} dx_2[n_\omega(x_0,x_2,x,\bar{x})+n_\omega(x_2,x_1,x,
\bar{x})].}}
The above equation can be further simplified by taking a Fourier transform with
respect to the
center of mass $\bar x$,
$$n_\omega(x_0,x_1,x,q)=\int d\bar{x}e^{iq\bar{x}}n_\omega(x_0,x_1,x,\bar{x})
.$$

 One of our key results is an exact solution to  \ntrans,
\eqn\solv{\eqalign{e^{-iq\bar{x}_{10}}n_\omega(x_0,x_1,x,q)=&{\delta(|x|-
x_{10})\over
\omega+\lambda x_{10}}+\theta(x_{10}-|x|)[{4\lambda^2\over(\omega+\lambda
|x|)^3q}\sin \left({x_{10}-|x|\over 2}q\right)\cr
&+{2\lambda\over(\omega+
\lambda |x|)^2}\cos\left({x_{10}-|x|\over 2}q\right)].}}
This solution can be understood intuitively. The first term is due to the
dipole formed by valence
quarks, the parent  dipole. The second term
comes from dipoles formed by emitted gluons. In particular,  the step function
in
the second term tells us that the size of induced dipoles cannot be greater
than $x_{10}$.
This is a direct consequence of the fact that no gluon is emitted outside
of the parent dipole. (We stress that $n_\omega(x_0,x_1,x,q)$ is not to be
confused with a dipole
density at a given transverse momentum $q$. Indeed, it is not even positive
definite.)

Transforming $n_\omega$ in \solv\ back to the rapidity variable, one finds
\eqn\momen{\eqalign{&e^{-iq\bar{x}_{10}}n(x_0,x_1,x,q,y)=\delta(|x|-x_{10})
e^{-\lambda x_{10}y}+\cr
&\theta(x_{10}-|x|)e^{-\lambda |x|y}\left({4\lambda^2y^2\over q}\sin
({x_{10}-|x|\over 2}q)+2\lambda y\cos({x_{10}-|x|\over 2}q)\right).}}
Surprisingly, the transform of  \momen\ back into the
center of mass coordinate takes on a very simple form:
\eqn\rapid{\eqalign{&n(x_0,x_1,x,\bar{x},y)=\delta(|x|-x_{10})\delta(\bar{x}
-\bar{x}_{10})e^{-\lambda x_{10}y}+\theta(x_{10}-|x|)e^{-\lambda|x|y}[
\lambda^2 y^2\cr
&\left(\epsilon(x_1-\bar{x}-|x|/2)-\epsilon(x_0-\bar{x}
+|x|/2)\right)+\lambda y\left(\delta(x_1-\bar{x}-|x|/2)+\delta(x_0-\bar{x}+
|x|/2)\right)].}}
It is easy to see that this expression for the  density is positive definite,
as it should.

\subsec{Meson-meson non-forward scattering}

Consider meson $A$ collides with meson $B$, in their center of mass system.
Denote the wave function of meson $A$ by $\Psi_A(x_{10},z)\exp(ip\bar{x}_{10})$
and the wave function of meson $B$ by $\Psi_B(x'_{10},z')\exp(-ip\bar{x}'_{10})
$. After integrating out their center of mass positions,  the scattering
amplitude between these two mesons can be expressed as
\eqn\mesc{A(s,t)=is\int dx_{10}dx'_{10}\Phi_A(x_{10})\Phi_B(x'_{10})A(x_{10},
x'_{10},Y,q),}
where $\Phi_A(x_{10})=\int dz|\Psi_A(x_{10},z)|^2$, $\Phi_B(x'_{10})=
\int dz'|\Psi_B(x'_{10},z')|^2$, and $A(x_{10},x'_{10},Y,q)$ is a scattering
amplitude between two
mesons with fixed dipole sizes. Under a two gluon exchange approximation, it
becomes
\eqn\disc{\eqalign{A(x_{10},x'_{10},Y,q)=&8g^4\int{dk\over 2\pi}
{1\over k^2(q-k)^2}
\int dxdx'n(x_0,x_1,x,q,Y/2)n(x'_0,x'_1,x',-q,Y/2)\cr
&\sin(kx/2)\sin((q-k)x/2)\sin(kx'/2)\sin((q-k)x'/2)),}}
where $Y=\ln s$, and $t=-q^2$.
Thus, the amplitude given
in Eq. \mesc\ can be associated with a (hard) Pomeron exchange. Indeed
as we shall see shortly, the leading contribution to $A(s,t)$ in Eq. \mesc\
comes from a
logarithmic branch point in the complex angular momentum plane.

Let us begin by first evaluating the following integral:
$$\int dxn(x_0,x_1,x,q,Y/2)\sin(kx/2)\sin((q-k)x/2).$$
We first point out that this quantity depends on the center of mass position
$\bar{x}_{10}$
only through the factor $\exp(iq\bar{x}_{10})$ as one can easily seen
from formula \momen. However, this dependence, together the similar factor
arising
from the second dipole density in \disc, will be removed when the
integration over center of mass positions is carried out in arriving at
\mesc. (The net result is a momentum conservation
factor which has been dropped from our formula \mesc.) We henceforth
ignore the dependence on $\bar{x}_{10}$ in the dipole density.
Next observe that, when $Y$ large,  the leading term in \momen\ is proportional
to $Y^2$. It follows that
\eqn\inte{\eqalign{&\int dxn(x_0,x_1,x,q,Y/2)\sin(kx/2)\sin((q-k)x/2)=
2\lambda^2Y^2\cr
&{k(q-k)\over q}\Im\left(e^{iqx_{10}/2}{\lambda Y+iq\over (\lambda^2Y^2
+2i\lambda qY)(\lambda^2Y^2+2i\lambda q Y-4k(q-k))}\right),}}
where $\Im(F)$ denotes the imaginary part.

In arriving at \inte,  we have initially assumed $x_1>x_0$.
However, it is easy to see \inte\ depends only on $|x_{10}|$. It follows that
the corresponding
integral in \disc\ involving the second dipole density can be  obtained
directly from
\inte\ by letting $q\rightarrow -q$, $k\rightarrow -k$ and $x\rightarrow
x'$. Substituting these results  into \disc\ and
performing the integral over $k$, we find
\eqn\what{\eqalign{A(x_{10},x'_{10},Y)=&-g^4{1\over q^2}({2\lambda Y
\cos(|x_{10}|-|x'_{10}|)q/2)\over \lambda^2 Y^2+4q^2}\cr
&-\lambda^2Y^2[
{\exp(iq(|x_{10}|+|x'_{10}|)/2)\over(\lambda Y+2iq)^2(\lambda Y+iq)}+
\hbox{c.c.}]).}}
This amplitude is non-singular in the limit $q\rightarrow 0$. When $|q|<<
\lambda Y$, it becomes
\eqn\becom{A(x_{10},x'_{10},Y)={4g^4\over\lambda q^2Y}\sin(qx_{10}/2)
\sin(qx'_{10}/2),}
and,  in the limit $q=0$, it behaves as
\eqn\forw{A(x_{10},x'_{10},Y)={g^4|x_{10}|x'_{10}|\over \lambda Y}.}

We point out that the amplitude decreases as $1/Y$ at high energies. This
result is in agreement
with a calculation done in Ref. \lt, using an effective action approach. Note
in the dipole picture, the important ingredient of virtual correction is
derived rather indirectly. In \lt, however, the decrease of scattering
amplitude is calculated with an effective action without additional input.
\becom\  can
also interpreted as  composed of a Pomeron ``propagator'' $1/(\lambda
Y)$ and its coupling to dipoles $x_{10}$ and $x'_{10}$. In this language, the
coupling
of a BFKL Pomeron to a dipole of size $x_{10}$ is  given by
$$\beta(q)={{2}g^2\over q}\sin(qx_{10}/2).$$

Finally substituting Eq.\what\ back to \mesc, one arrives at
\eqn\ampl{A(s,t)=-isg^4{4\over q^2}[{\lambda Y\Phi_A(q/2)\Phi_B(-q/2)\over
\lambda^2Y^2+4q^2}-{\lambda^2Y^2\Phi_A(q/2)\Phi_B(q/2)\over(\lambda Y+
2iq)^2(\lambda Y+iq)}+\hbox{c.c}],}
where
$$\Phi_A(q/2)=\int_0^\infty dx\Phi_A(x)e^{iqx/2},\quad
\Phi_B(q/2)=\int_0^\infty dx\Phi_B(x)e^{iqx/2}$$
and $\Phi_A(-q/2)$, $\Phi_B(-q/2)$ are their complex conjugates. (We assume
that both $\Phi_A(x)$ and $\Phi_B(x)$ are even functions.)  For the forward
limit $q=0$, one can use \forw\ with $|x_{10}|$ and $|x'_{10}|$ replaced
by the effective sizes of meson $A$ and meson $B$. Note Also that the high
energy behavior
\forw\ can be understood as due to  a logarithmic branch point in the complex
angular momentum plane, {\it i.e.},
$$A(x_{10},x'_{10},Y)=-{g^4|x_{10}||x'_{10}|\over 4\lambda}\int{d\omega
\over 2\pi i}s^{\omega}\ln (\omega).$$
The branch point is located at $\omega=0$.

Physically, the asymptotic decrease of the scattering amplitude is due to a
suppression
of the dipole density at large rapidity. This can be understood as follows.
For 2+1 dimensions,
since there is only one transverse dimension, the emitted gluons in a meson is
confined
between the valence quarks, and the branching process of emission of soft
gluons must compete  with a suppression factor $\exp(-\lambda|x|y)$ due to
virtual corrections.  In 3+1
dimensions, this competition is won by the branching process, hence a
BFKL Pomeron emerges to the right of $J=1$ in the complex angular momentum
plane. In 2+1 dimensions,
however, the branching process losses the competition and the suppressing
factor due to  virtual
corrections  dominates, thus leading to a logarithmic branch point at $J=1$.

\newsec{Two Pomeron exchange}

\subsec{Dipole pair density}

The next level contribution to the scattering amplitude in the leading
logarithmic approximation is from exchanges of four gluons between two
mesons. In order to calculate this contribution, we need to
know the density of two dipoles in a given meson state. Let $n_2(x_0,x_1,
x_a,\bar{x}_a,x_b,\bar{x}_b,y)$ be the density of two dipoles with respective
center
of mass positions $\bar{x}_a$ and $\bar{x}_b$ and separations $x_a$ and
$x_b$. Again this density is defined for dipoles with lowest longitudinal
momentum greater than $e^{-y}p_+$, where $p_+$ is the meson longitudinal
momentum.
(In deriving an integral equation for $n_2$, we will first  focus on its
dependences on $x_0, x_1$
and $y$. In the following, for compactness, we shall sometimes drop its
dependences on
$x_a,\bar{x}_a,x_b$, and $\bar{x}_b$. They will be re-inserted when necessary.)

After a careful
calculation, we have found that the leading term in the double density is
simply the product of
two single dipole densities. This will be demonstrated below. This surprising
result implies that,
to the leading order, there is no correlation between two single dipole
densities. This is not the
case for 3+1 dimensions. We take this as one of the remarkable features of the
Pomeron physics in
three dimensions.

The integral equation can be written down readily by examining the graphical
equation in
figure 6:
\eqn\ddip{\eqalign{n_2(x_0,x_1,y)=&\lambda\int_{x_0}^{x_1}dx_2\int_0^y
e^{-\lambda x_{10}
(y-y')}dy'[n(x_2,x_1,x_a,\bar{x}_a,y')n(x_0,x_2,x_b,\bar{x}_b,y')\cr
&+n(x_2,x_1,x_b,\bar{x}_b,y')n(x_0,x_2,x_a,\bar{x}_a,y')
+n_2(x_2,x_1,y')+n_2(x_0,x_2,y')],}}
where again we assumed $x_1>x_0$. The first two terms come from finding two
dipoles in separated dipoles, as indicated in figure 6, which serve as
the inhomogeneous term in the integral equation. Note that the single dipole
density depends on the separation of the dipole only through its absolute
value, the same is true for the double dipole density, as evident from the
above equation.
\vskip1cm
\epsfxsize=340pt \epsfbox{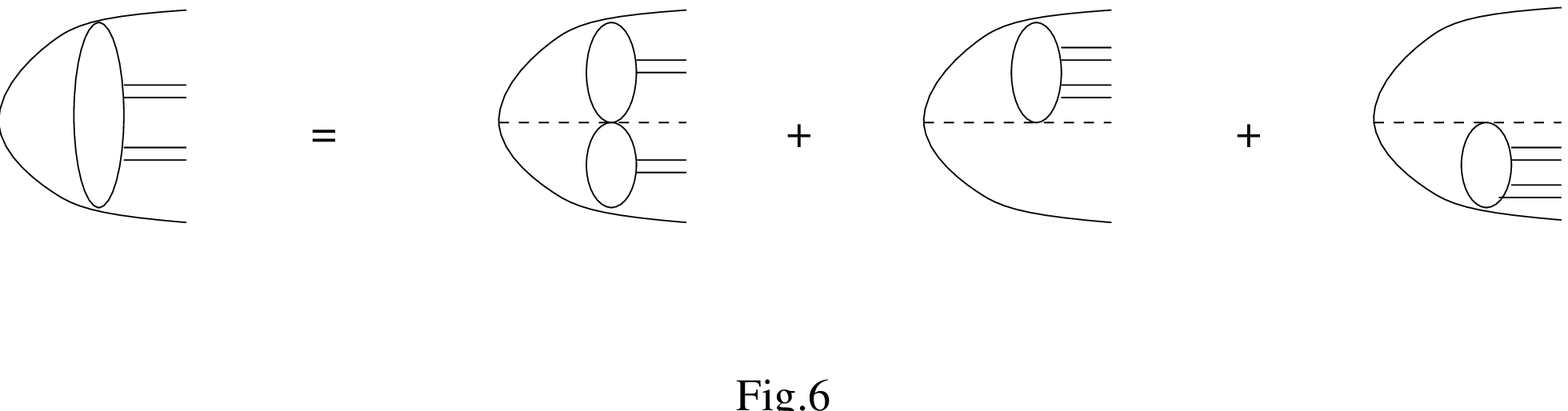}

The single dipole density,
as given in \rapid, is already lengthy. Fortunately, in order to calculate  the
leading term in the double dipole density,  it is enough to only keep
the leading term in the single dipole density as the input. This term,
from \rapid, is
\eqn\leading{\eqalign{n(x_0,x_1,x,\bar{x},y)=&\lambda^2y^2e^{-\lambda xy}
\theta(x_{10}-x)(\epsilon(x_1-\bar{x} -x/2)-\epsilon(x_0-\bar{x}+x/2))\cr
&=2\lambda^2y^2e^{-\lambda xy}\theta(x_{10}-x)\theta(x_1-\bar{x}-x/2)\theta
(\bar{x}-x_0-x/2).}}
Plugging this into \ddip, the integral equation reduces to
\eqn\tos{\eqalign{n_2(x_0,x_1,y)=&n_2^{(0)}(x_0,x_1,y)\cr
&+\lambda\int_{x_0}^{x_1}dx_2\int_0^ydy'e^{-\lambda
x_{10}(y-y')}
\left(n_2(x_2,x_1,y')+n_2(x_0,x_2,y')\right).}}
where
\eqn\toso{\eqalign{n_2^{(0)}(x_0,x_1,y)&=4\lambda^5e^{-\lambda x_{10}y}\int_0^y
y'^4
e^{\lambda(x_{10}-x_a-x_b)y'}dy'[
(\bar{x}_a-\bar{x}_b-(x_a+x_b)/2)\cr
&\theta
(\bar{x}_a-\bar{x}_b-(x_a+x_b)/2)
\theta(x_1-\bar{x}_a-x_a/2)\theta(\bar{x}_b-
x_0-x_b/2)+(a\leftrightarrow b)].}}
The presence of various step functions assures that both dipoles are  found
only within the parent
dipole, and they do not overlap.

Eq. \tos\ can be solved by taking a Laplace transform with respect to $y$, and
this is carried  in
the Appendix. The leading order term in $y$ takes on a surprisingly simple form
\eqn\final{\eqalign{n_2(x_0,x_1,
x_a,\bar{x}_a,x_b,\bar{x}_b,y)=&4\lambda^4y^4e^{-\lambda(x_a+x_b)y}
\theta(\bar{x}_a-\bar{x}_b-(x_a+x_b)/2)\cr
&\theta(x_1-\bar{x}_a-x_a/2)\theta(\bar{x}_b-x_0-x_b/2)+(a\leftrightarrow
b).}}
Comparing this result to the single dipole density in Eq. \leading, we find
that the leading term of the double dipole density, as given in
\final, is just the product of two single dipole densities, the only additional
requirement is that the two dipoles do not overlap. The first
term corresponds to the situation that the center
of mass position of dipole $a$ is above the center of mass position of
dipole $b$. The second term represents the other possibility. This result
is not at all obvious before we solve the integral equation \tos. Note
that the inhomogeneous term in \tos, given in the first graph on the r.h.s.
of figure 6, is different from \final, since its leading term in $y$
is proportional to $\bar{x}_a-\bar{x}_b-(x_a+x_b)/2$, which is absent
in \final. With this factor, the density
is not homogeneous regarding to the center of mass positions. The result
in \final\ is rather homogeneous in the center of mass positions.

\subsec{Meson-meson amplitude due to two Pomeron exchange}

With formula \final\ of the dipole pair density, we can calculate the
scattering amplitude due to  four gluon exchanges between a pair
of dipoles in one meson and another pair of dipoles in the other meson.
The scattering amplitude with zero momentum transfer, for given parent
dipole sizes $x_{10}$ and $x'_{10}$, is given by
\eqn\twpair{\eqalign{&A_2(x_{10}x'_{10},q=0,Y)={2!\over 2!\times 2!}
\int{dk\over 2\pi}dx_adx_bdx'_adx'_bn_2(x_0,x_1,x_a,k,x_b,-k, Y/2)\cr
&n_2(x'_0,x'_1,x'_a,k,x'_b,-k,Y/2)[8g^4{dk_a\over 2\pi k_a^2(k-k_a)^2}
d(k_a,k,x_a)d(k_a,k,x'_a)]\cr
&[8g^4{dk_b\over 2\pi k_b^2(k+k_b)^2}d(k_b,-k,x_b)d(k_b,-k,x'_b)],}}
where the term in the first bracket is from the exchange of two gluons between
dipoles $x_a$ and $x'_a$, and the term in the second bracket is from
the exchange of two gluons between dipoles $x_b$ and $x'_b$. In \twpair,
$d(k_a,k,x_a)=\sin(k_ax_a/2)\sin((k-k_a)x_a/2),$
 and the dipole density
$n_2(x_0,x_1,x_a,k,x_b,-k,Y/2)$
is the Fourier transform of
$n_2(x_0,x_1,x_a,\bar{x}_a,x_b,\bar{x}_b,Y/2)$
with respect to the
center of mass positions, (relative to the center of mass of the parent
dipole).  Finally, the numerical factor $2!$ is
from possible pairing among two pairs of dipoles, and two factors $2!$ in
the denominator are due to indistinquishability of two dipoles in a meson.
A version of Eq.\twpair\ in two transverse dimensions has been derived in the
Appendix of  \mp.

Were the dipole density $n_2$ more complicated than given in Eq. \final,
it would be a hard task to perform the integral in \twpair. Fortunately, Eq.
\final\
tells us that in the leading order, the dipole pair density is the
product of two single dipole densities, up to the requirement that
the two dipoles do not overlap. Recall in our calculation of the amplitude
due to one Pomeron exchange, the major contribution is from dipole size
$x\sim 1/(\lambda Y)$, which is very small at high energies. The smallness
of dipole sizes then allows us to neglect the requirement that two dipoles
do not overlap in a given meson, then the dipole pair density is simply
the product of two single dipole densities. The integral in \twpair\
factorizes. Thus, the two Pomeron exchange amplitude is given by
\eqn\simp{\eqalign{A_2(x_{10},x'_{10},q=0,Y)=&{1\over 2}\int{dk\over 2\pi}
A(x_{10},x'_{10},k,Y)A(x_{10},x'_{10},-k,Y)\cr
=&{1\over 2}\int{dk\over 2\pi}A^2(x_{10},x'_{10},k,Y),}}
where the last equality derives from the fact that the single Pomeron
exchange amplitude, as given in \what, is an even function of $k$. The
above formula allows us the calculate the forward scattering amplitude
$A_2$ from the knowledge of $A$. Performing the integral over $k$ in
\simp\ and keeping only the leading term, we have
\eqn\twpom{A_2(x_{10},x'_{10},q=0,Y)={g^8\over 6\lambda^2Y^2}\left({1\over 2}
|x_{10}-x'_{10}|^3+{1\over 2}(x_{10}+x'_{10})^3-x^3_{10}-(x'_{10})^3
\right).}
So the amplitude is suppressed by the factor $1/(\ln s)^2$, comparable to
the sub-leading term in the single Pomeron exchange amplitude. The dependence
of $A_2$ on the dipole sizes is cubic, as the correct dimensionality
requires. In principle one can also compute $A_2$ for a finite momentum
transfer. The formula is given by
\eqn\finitr{A_2(x_{10},x'_{10},q,Y)={1\over 2}\int {dk\over 2\pi}A(x_{10},
x'_{10},k,Y)A(x_{10},x'_{10},q-k,Y).}

\newsec{The triple Pomeron coupling}

We have found a simple formula for the scattering amplitude of two Pomeron
exchange, indicating the simplicity of Pomeron physics in the three
dimensional QCD. The next important object to calculate is the triple Pomeron
coupling. Armed with knowledge of all these, one may attempt to construct
the Regge field theory which helps one to study high energy scattering
systematically. It is interesting, as one may expect, that once again
the triple Pomeron coupling can be calculated exactly to the leading order,
unlike the situation in the four dimensional QCD.

The physical process involving triple Pomeron coupling is $A+B\rightarrow
C+X$, where $A$, $B$ and $C$ are mesons and $X$ is anything. The
generalized optical theorem states that the partial cross section
of this so-called one-particle inclusive process
is given by the imaginary part of amplitude of the elastic process $A+B+\bar{C}
\rightarrow A+B+\bar{C}$ \ref\hz{see for example, D. Horn and F. Zachariasen,
Hadron
Physics at Very High Energies, W.A. Benjamin, Inc. 1973.}. To be more precise,
let $p$ be the out-going momentum of $C$, then $E_Cd\sigma/d^2p={1\over 2
s(2\pi)^2}\Im (A_6) $, where $A_6$ is the connected part of the forward
scattering amplitude of $A+B+\bar{C}\rightarrow A+B+\bar{C}$. To extract
information concerning the triple Pomeron coupling, we consider the case
$C=A$ and the one particle inclusive process is dominated by a Pomeron
exchange. Instead of considering some definite meson states, let us consider
mesons with a fixed dipole size. Let $x_{10}$ the dipole size of the $B$
meson and $x'_{10}$ the dipole size of the $A$ meson. Furthermore, we
define $M^2$ be the mass squared of $X$, rapidity $\bar{y}=\ln ({s}/M^2)$,
where $s$ is the total energy squared. Also define $Y=\ln({s}/m_T^2)$, $m_T$
is the transverse mass of the outgoing $A$ meson. With these notations,
it is easy to see that
$$E_C{d\sigma\over d^2p}=s{d\sigma\over dqdM^2},$$
where $q$ is the momentum transfer. Consider large $\bar{y}$, the process
is then dominated by a single Pomeron exchange, so
$$s{d\sigma\over dq^2dM^2}\sim {1\over s}{g^4\over q^2}\left(\sin(qx'_{10}/2)
\right)^2 {4\over \lambda^2\bar{y}^2}|\sum_X\beta_{BX}|^2$$
where we have used the Pomeron ``propagator'' and its coupling to meson
$A$, $\beta_{BX}$ denotes the Pomeron coupling to $B$ and $X$.
Again applying the optical theorem, $|\sum_X\beta_{BX}|^2$ can be written
as a product of the triple coupling coupling, the Pomeron propagator
at $Y-\bar{y}=\ln(M^2/m_T^2)$ and the Pomeron coupling to meson $B$. Note
that this coupling to $B$ is the one at zero momentum transfer. Putting
everything together, one has
\eqn\tric{E_C{d\sigma\over d^2p}\sim{M^2\over s}{8g^6x_{10}\over
\lambda^3\bar{y}^2(Y-\bar{y})|q|}\left(\sin(qx'_{10}/2)\right)^2V_{3P}(q),}
where $V_{3P}$ shall be referred to as  the ``triple Pomeron coupling".

Next, we calculate the 6-point amplitude $A_6$ and use the optical theorem
to relate this amplitude to the above quantity, and finally read off
$V_{3P}$. One can imagine that such amplitude is due to one Pomeron exchange
between $A$ and $B$ and one Pomeron exchange between another $A$ and $B$.
So for meson $B$ we need a dipole pair density. The rapidity available for
one Pomeron exchange is $\bar{y}$. However for the dipole pair in meson $B$,
the rapidity available is different from $\bar{y}$, and $\bar{y}$ serves only
as a rapidity cut. We then follow \mp\ to define the dipole pair density
depending on both $Y$ and $\bar{y}$, here $Y$ is the maximum rapidity.

The desired equation again can be readily read off from a graphical equation,
which we spare here:
\eqn\dfev{\eqalign{&n_2(x_0,x_1,y,\bar{y})=\lambda\int_{x_0}^{x_1}
dx_2e^{-\lambda x_{10}(y-\bar{y})}[n(x_2,x_1,x_a,\bar{x}_a,\bar{y})
n(x_0,x_2,x_b,\bar{x}_b,\bar{y})\cr
&+(a\leftrightarrow b)]+\lambda\int_{x_0}
^{x_1}dx_2\int^y_{\bar{y}}dy'e^{-\lambda x_{10}(y-y')}[n_2(x_2,x_1,y',\bar{ y})
+n_2(x_0,x_2,y',\bar{y})],}}
where all quantities with subscript $a$ are those associated to meson $A$,
and those with subscript $b$ are associated to meson $B$.
The solution to this equation can be obtained in a similar way as we solved
Eq.\ddip. We skip the details and directly write down the answer
\eqn\final{\eqalign{n_2(x_0,x_1,y,\bar{y})=&4\lambda^5\bar{y}^4e^{-
\lambda(\bar{x}_a-\bar{x}_b)(y-\bar{y})-\lambda(x_a+x_b)(y+\bar{y})/2}(
\bar{x}_a-\bar{x}_b-(x_a+x_b)/2)\cr
&\theta(\bar{x}_a-\bar{x}_b-(x_a+x_b)/2)
\theta(x_1-\bar{x}_a-x_a/2)\theta(\bar{x}_b-x_0-x_b/2)+(a\leftrightarrow
b).}}
We remark that this solution is exact with the input of the single dipole
density given in \leading.
Next we need to transform this density into the momentum space, which is
the quantity needed in calculate the 6-point amplitude. Assuming $q=q_a
=-q_b$ and ignoring the dipole sizes in various theta functions, we find
\eqn\dimom{\eqalign{&n_2(x_0,x_1,q,-q,y,\bar{y})=\int d\bar{x}_ad\bar{x}_b
e^{iq(\bar{x}_a-\bar{x}_b)}n_2(x_0,x_1,\bar{x}_a,\bar{x}_b,y,\bar{y})\cr
&=4\lambda^5\bar{y}^4
e^{-\lambda(x_a+x_b)(y+\bar{y})/2}[{x_{10}
\over(\lambda(y-\bar{y})+iq)^2}(e^{-(\lambda(y-\bar{y})+iq)x_{10}}+1)\cr
&+{2\over(\lambda(y-\bar{y})+iq)^3}(e^{-(\lambda(y-\bar{y})+iq)x_{10}}-1)+
\hbox{c.c.}].}}
This formula is still a bit too long. Indeed the physical situation
corresponds to $\lambda(y-\bar{y})x_{10}>>1$, and this simplifies the above
formula a lot:
\eqn\simm{n_2(x_0,x_1,q,-q,y,\bar{y})=8\lambda^5\bar{y}^4x_{10}{\lambda^2(y-
\bar{y})^2-q^2\over(\lambda^2(y-\bar{y})^2+q^2)^2}e^{-\lambda(x_a+x_b)(y+
\bar{y})/2}.}

Now we are ready to calculate the 6-point amplitude. By definition, this
amplitude is given by
\eqn\twpai{\eqalign{&A_6(q)=iM^2
\int dx_adx_bdx'_adx'_bn_2(x_0,x_1,x_a,q,x_b,-q,Y-{\bar{y}\over
2},{\bar{y}\over 2})
n(x'_{10},x'_a,q,{\bar{y}\over 2})\cr
&n(x'_{10},x'_b,q,{\bar{y}\over 2})[8g^4{dk_a\over 2\pi k_a^2(q-k_a)^2}
d(k_a,k,x_a)d(k_a,k,x'_a)]\cr
&[8g^4{dk_b\over 2\pi k_b^2(q-k_b)^2}d(k_b,-k,x_b)d(k_b,-k,x'_b)],}}
where $Y-\bar{y}=\ln M^2$ and  other notations are the same as in the previous
section.
Recall that the single dipole density in the momentum space is
$$n(x'_{10},x'_a,q,\bar{y})=4\lambda^2\bar{y}^2{1\over q}\sin (qx'_{10}/2)
e^{-\lambda x'_a\bar{y}},$$
again we ignored the dipole separation as compared to $x'_{10}$. Substituting
this formula and the formula in Eq.\simm\ into \twpai, we find
$$\eqalign{&A_6(q)={i\over 2} \lambda^9M^2\bar{y}^8{\lambda^2(Y-\bar{y})^2-q^2
\over (\lambda^2(Y-\bar{y})^2+q^2)^2}{x_{10}\over q^2}\left(\sin(qx'_{10}/2)
\right)^2\cr
&\left(8g^4\int dx_adx_a'e^{-\lambda x_aY/2-\lambda
x_a'\bar{y}/2}{dk_a\over 2\pi
k_a^2(q-k_a)^2}d(k_a,q,x_a)d(k_a,q,x'_a)\right)^2.}
$$
Finally performing the integral in the above equation and one is led to a
quite lengthy formula
\eqn\six{\eqalign{A_6(q)=&2i g^8M^2\lambda^7\bar{y}^8{x_{10}\over (Y+
\bar{y})^2q^2}
\left(\sin(qx'_{10}/2)\right)^2\cr
&{\lambda^2(Y-\bar{y})^2-q^2\over(\lambda^2(Y-\bar{y})^2+q^2)^2(\lambda^2
Y^2+q^2)^2(\lambda^2\bar{y}^2+q^2)^2}.}}
This formula is actually not valid when $q\sim\lambda Y$, since we have ignored
the
dipole size in the sine factors in the single dipole density as well as in
the dipole pair density. This is to be contrasted with the case for one Pomeron
exchange amplitude,
 \what, where the dependence on dipole
sizes
is not ignored. (If one so desires, one can of course take the dependence on
dipole sizes  into account, leading to a much more complicated function of
$q$.)  When $|q|<<\lambda Y$ where  Eq. \six\ is valid, it simplifies to
\eqn\simp{A_6(q)=2ig^8M^2{\bar{y}^4x_{10}\over \lambda^3Y^4(Y+
\bar{y})^2(Y-\bar{y})
^2}{1\over q^2}\left(\sin(qx'_{10}/2)\right)^2.}
Upon using the optical theorem, $E_cd\sigma/d^2p={1\over 2s(2\pi)^2}\Im (A_6)$,
and comparing  \simp\ to \tric, we find that the triple
Pomeron coupling depends on rapidities:
\eqn\tricp{V_{3P}\sim {g^2\over |q|}{\bar{y}^6\over Y^4(Y+\bar{y}
)^2(Y-\bar{y})}.}
It tends to zero when both $Y$ and $\bar{y}$ are large. It is also singular
when $\sqrt{-t}=|q|\rightarrow 0$, just as in the 3+1 dimensional case \mp.
However the origin of this singularity in the present case is purely
kinematic, as Eq. \tric\ shows. If the one particle inclusive cross section
factorizes, one expects that $V_{3P}$ is independent of rapidities. Our result
indicates that the inclusive cross section does not factorize and
the triple Pomeron coupling  vanishes as  $Y^{-1}$, with $Y-\bar{y}$ fixed.

\newsec{Discussion}

As we have seen, one remarkable property of high energy scattering processes
in our toy world, 2+1 dimensional QCD, is the fact that no soft on-shell gluon
can
escape from hadron emitting it. This is due to the quantum interference
of quark and anti-quark. This property is also the reason underlying the fact
that the total cross section decreases at high energies, as was first pointed
out in \lt. It  is also responsible  for the interesting fact that, in the
leading order, there is no
correlation  between a pair of dipoles in a meson state, so the
amplitude due to two Pomeron exchange is simply given by Eq. \finitr.
Physically, one expects that this holds true even for multi-dipole density.
Thus, in the leading order, the amplitude due to n-Pomeron exchange is given
by
\eqn\npom{A_n(x_{10},x'_{10},q)={1\over n!}\int\prod_{i=1}^{n-1}\left(
{dk_i\over 2\pi}A(x_{10},x'_{10},k_i)\right)A(x_{10},x'_{10},q-\sum_ik_i),}
where $A(x_{10},x'_{10},k)$ is the amplitude due one Pomeron exchange between
dipole $x_{10}$ and dipole $x'_{10}$. \npom\ is a straightforward
generalization of  \finitr. Going to the impact parameter space, Eq. \npom\
implies that the eikonal formula for summing over all possible number
of  Pomeron exchanges is valid in the leading order. However, as can be seen
in Sec. 3, there are sub-leading terms on each level of Pomeron exchanges,
and these terms
are comparable to the leading term on the next level of Pomeron exchanges.

Of course, to incorporate the
whole complexity of high energy scattering, one needs to include various
couplings among Pomerons. The triple Pomeron coupling is already calculated
in this work. It is interesting to see whether one can develop a Reggeon
field theory based on our results.
\vfill
\noindent{\bf Acknowledgments}

We would like to thank E. M. Levin, A.H. Mueller and A.R. White for
interesting discussions on high energy
hadronic collision in the near-forward limit. This work was supported
by DOE contract DE-FG02-91ER40688-Task A.
\vfill\eject
\noindent{\bf Appendix}

Eq. \tos\ can be simplified by a Laplace transform.  Denoting the transform for
$n_2$ by $n_{2,\omega}(x_0,x_1)$, where the variable $\omega$ plays the role of
a ``complex angular
momentum ", the integral over $y'$
can be undone. Let
$$\tilde{n}_{2,\omega}(x_0,x_1)={1\over 4\times 4!}\left(\omega+\lambda(x_a+
x_b)\right)^5n_{2,\omega}(x_0,x_1),$$
we obtain the integral equation for $\tilde{n}_{2,\omega}$:
$${\eqalign{&(\omega+\lambda x_{10})\tilde{n}_{2,\omega}(x_0,x_1)=
\lambda^5[(\bar{x}_a-\bar{x}_b-(x_a+x_b)/2)\theta(\bar{x}_a-\bar{x}_b-(x_a+
x_b)/2)\cr
&\theta(x_1-\bar{x}_a-x_a/2)
\theta(\bar{x}_b-x_0-x_b/2)+(a\leftrightarrow b)]
+\lambda\int_{x_0}^{x_1}dx_2\left(\tilde{n}_{2,\omega}(x_2,x_1)+\tilde{n}
_{2,\omega}(x_0,x_1)\right).}}\eqno (A.1)$$
To solve (A.1), we use the following ansatz
$${\eqalign{\tilde{n}_{2,\omega}(x_0,x_1)=&\lambda^5(\bar{x}_a-
\bar{x}_b-(x_a
+x_b)/2)\theta(\bar{x}_a-\bar{x}_b-(x_a+x_b)/2)\theta(x_1-\bar{x}_a-x_a/2)\cr
&\theta(\bar{x}_b-x_0-x_b/2)f_\omega(x_0,x_1,x_a,\bar{x}_a,x_b,\bar{x}_b)
+(a\leftrightarrow b).}}\eqno (A.2)$$
The unknown function $f_\omega$ then satisfies
$$(\omega+\lambda x_{10})f_\omega(x_0,x_1)=1+\lambda[\int_{x_0}^{\bar{x}_b-
x_b/2}dx_2f_\omega(x_2,x_1)+\int_{\bar{x}_a+x_a/2}^{x_1}dx_2f_\omega (x_0,
x_2)].$$
 The solution is
$$f_\omega=[\omega+\lambda(\bar{x}_a-\bar{x}_b+(x_a+x_b)/2)]^{-1},$$
which is independent of
$x_0$ and $x_1$.   Substituting this result  in (A.2) and recalling the
definition for
$\tilde{n}_{2,\omega}$, we find
$${\eqalign{&n_{2,\omega}(x_0,x_1)=4\times 4!\lambda^5[\omega+\lambda
(x_a+x_b)]^{-5}[\omega+\lambda(\bar{x}_a-\bar{x}_b+(x_a+x_b)/2)]^{-1}(\bar{x}_a
-\bar{x}_b\cr
&-(x_a+x_b)/2)
\theta(\bar{x}_a-\bar{x}_b-(x_a+x_b)/2)\theta(x_1-\bar{x}_a-x_a/2)\theta
(\bar{x}_b-x_0-x_b/2)+(a\leftrightarrow b).}}\eqno (A.3)$$
It is easy to transform $n_{2,\omega}$
back to arrive at $n_2(x_0,x_1,y)$. There are many terms each proportional to a
certain
power of $y$. The leading term takes on the following simple form:
$${\eqalign{n_2(x_0,x_1,
x_a,\bar{x}_a,x_b,\bar{x}_b,y)=&4\lambda^4y^4e^{-\lambda(x_a+x_b)y}
\theta(\bar{x}_a-\bar{x}_b-(x_a+x_b)/2)\cr
&\theta(x_1-\bar{x}_a-x_a/2)\theta(\bar{x}_b-x_0-x_b/2)+(a\leftrightarrow
b).}}\eqno (A.4)$$
\vfill\eject
\listrefs

\centerline{\bf Figure captions}

\noindent Fig.1:  A soft gluon emitted either from the quark or the anti-quark
of a dipole. The dashed line denotes the gluon.

\noindent Fig.2: A graphical  illustration of Eq. (2.6) for the generating
functional
$Z$.

\noindent Fig.3: A string picture of meson-meson scattering by a closed string,
a Pomeron, exchange.

\noindent Fig.4: Quark-gluon diagram of the process of two gluon exchange
and its corresponding surface picture, obtained by replacing a gluon line
by two quark lines.

\noindent Fig.5: A graphical  equation for the dipole density, again dashed
lines
are soft gluon lines.

\noindent Fig.6: A graphical equation for the dipole pair density.

\vfill\eject

\end